\documentclass[sigconf]{acmart}
 
\usepackage[framemethod=tikz]{mdframed}
\AtBeginDocument{%
  \providecommand\BibTeX{{%
    \normalfont B\kern-0.5em{\scshape i\kern-0.25em b}\kern-0.8em\TeX}}}

\renewcommand\footnotetextcopyrightpermission[1]{} % removes footnote with conference information in first column
\usepackage{amsmath,amsfonts}
\usepackage{algorithmic}
\usepackage{graphicx}
\usepackage{textcomp}
\usepackage{multirow}
\usepackage{booktabs}
\usepackage{threeparttable}
\usepackage{enumitem}
\usepackage[framemethod=tikz]{mdframed}
\usepackage{tikz}
\usepackage{tipa}
\usepackage{xtab}
\usepackage{mfirstuc}
\usepackage{todonotes}
\usepackage{soul}
\usepackage{colortbl}
\usepackage{xspace}
\usepackage{xcolor}
\usepackage{url}

\usepackage{caption}
\usepackage{subcaption}
\usepackage{wrapfig}
\usepackage[many]{tcolorbox}
\newtcolorbox{mybox}[1][]{
  breakable,
  fonttitle=\bfseries,
  bottomrule=0pt,
  toprule=0pt,
  leftrule=1pt,
  rightrule=1pt,
  titlerule=0pt,
  arc=0pt,
  outer arc=0pt,
  colframe=black,
}

\newboolean{showcomments}
\setboolean{showcomments}{false} % toggle to show or hide comments
\ifthenelse{\boolean{showcomments}}
    {

        \newcommand{\zane}[1]{\todo[inline,linecolor=blue,backgroundcolor=blue!25,bordercolor=blue]{\textcolor{black}{\textbf{Zane says:}} #1}}
        \newcommand{\nowshin}[1]{\todo[inline,linecolor=red,backgroundcolor=red!25,bordercolor=red]{\textcolor{black}{\textbf{Nowshin says:}} #1}}
        \newcommand{\kezia}[1]{\todo[inline,linecolor=cyan,backgroundcolor=yellow!25,bordercolor=purple]{\textcolor{black}{\textbf{Kezia says:}} #1}}
        \newcommand{\manish}[1]{\todo[inline,linecolor=green,backgroundcolor=green!25,bordercolor=green]{\textcolor{black}{\textbf{Manish says:}} #1}}
        \newcommand{\dana}[1]{\todo[inline,linecolor=yellow,backgroundcolor=orange,bordercolor=black]{\textcolor{black}{\textbf{Dana says:}} #1}}
        \newcommand{\neil}[1]{\todo[inline,linecolor=yellow,backgroundcolor=yellow,bordercolor=black]{\textcolor{black}{\textbf{Neil says:}} #1}}
       
    }
    {
        \newcommand{\zane}[1]{}
        \newcommand{\nowshin}[1]{}
        \newcommand{\kezia}[1]{}
        \newcommand{\manish}[1]{}
        \newcommand{\dana}[1]{}
        \newcommand{\neil}[1]{}
    }

\newcommand{\q}[1]{\emph{``#1''}}
\newcommand{\qs}[2]{\emph{``#1'' (#2)}}

\newcommand{\org}[1]{\emph{(Orgs.: #1)}}

\renewcommand{\theenumii}{\theenumi.\textbf{\arabic{enumii}}}

\begin{document}

\title [Unveiling the Life Cycle of User Feedback: Best Practices from Software Practitioners]{Unveiling the Life Cycle of User Feedback: Best Practices from Software Practitioners} 
% \title [The Secret Life of User Feedback: How Organizations Manage It]{The Secret Life of User Feedback: How Organizations Manage It} 
% \title [Are you Listening? The missing piece to product innovation: Voices of Users]{Are you Listening? The missing piece to product innovation: Voices of Users} 
% \title [Are you listening? Becoming a high performing organization through leveraging the Voices of Users]{Are you listening? Becoming a high performing organization through leveraging the Voices of Users} 
% \title [mode = title]{Have you read this yet? The missing piece to product innovation: Voice of Users} 
%Unveiling the Life Cycle of User Feedback: Best Practices from Software Practitioners

%Authors
\author{Ze Shi Li}
\affiliation{%
  \institution{University of Victoria}
  \country{Victoria, Canada}
}
\email{lize@uvic.ca}

\author{Nowshin Nawar Arony}
\affiliation{%
  \institution{University of Victoria}
  \country{Victoria, Canada}
}
\email{nowshinarony@uvic.ca}

\author{Kezia Devathasan}
\affiliation{%
  \institution{University of Victoria}
  \country{Victoria, Canada}
}
\email{keziadevathasan@uvic.ca}

\author{Manish Sihag}
\affiliation{%
  \institution{University of Victoria}
  \country{Victoria, Canada}
}
\email{manishsihag@uvic.ca}

\author{Neil Ernst}
\affiliation{%
  \institution{University of Victoria}
  \country{Victoria, Canada}
}
\email{nernst@uvic.ca}

\author{Daniela  Damian}
\affiliation{%
  \institution{University of Victoria}
  \country{Victoria, Canada}
}
\email{danielad@uvic.ca}

%Abstract
\begin{abstract}
% Social media is becoming an increasingly popular place for users to provide feedback about software products. 
User feedback has grown in importance for organizations to improve software products. 
% User feedback and suggestions about software can quickly become trending topics, attracting widespread attention and discussion. 
% However, effectively managing user feedback, including social media-based feedback, presents significant challenges for software organizations, especially regarding identifying areas for improvement and incorporating feedback into development.
%However, recent studies analyzing industry practices primarily address the feedback collection process but lack details on how practitioners reason about and use this user feedback to improve their software, or gain competitiveness in the market.
Prior studies focused primarily on feedback collection and reported a high-level overview of the processes, often overlooking how practitioners reason about, and act upon this feedback through a structured set of activities.
In this work, we conducted an exploratory interview study with 40 practitioners from 32 organizations of various sizes and in several domains such as e-commerce, analytics, and gaming.
%, where we investigated how they engage in managing the user feedback.
%Our findings indicate that user feedback comes from a rich variety of sources, including social media, support channels, workshops, and user usage metrics, and that organizations are afforded unprecedented opportunities to monitor user perceptions and react timely to prevent risk of misinterpretation or loss of revenue. 
% Our findings indicate that user feedback originates from an abundance of different sources, including social media. Thus, organizations are afforded unprecedented opportunities to monitor rapidly evolving user perceptions and react in a timely manner to mitigate the risks of feedback misinterpretation. 
Our findings indicate that organizations leverage many different user feedback sources. 
Social media emerged as a key category of feedback that is increasingly critical for many organizations.   
% Although the participant organizations did not follow a systematic approach for managing user feedback, 
We found that organizations actively engage in a number of non-trivial activities to curate and act on user feedback, depending on its source. 
We synthesize these activities into a \emph{life cycle of managing user feedback.}
% For example, while social media feedback uniquely provides fast and ample insights on bugs in a product, it is also noisier, requiring ongoing monitoring and \emph{analysis} for developing trends, and \emph{validation} for accuracy in tandem with other sources of user feedback, before it is actionable for the \emph{prioritization} of features/bugs in development.
%Although, we did not derive a systematic approach for managing user feedback, we identify a number of non-trivial activities organizations use to effectively act on user feedback. 
% We synthesize a number of common activities that emerged across organizations into a . 
We also report on the \emph{best practices for managing user feedback} that we distilled from responses of practitioners who felt that their organization effectively understood and addressed their users’ feedback. 
%These practices include \emph{active employee participation}, \emph{active feedback collection}, \emph{active triangulation}, and \emph{active response to user feedback}. 
We present actionable empirical results that organizations can leverage to increase their understanding of user perception and behavior for better products thus reducing user attrition.

%Our participants highlighted the importance of considering user feedback from a variety of sources, including social media, support channels, workshops, and user usage metrics. Organizations are afforded unprecedented opportunities to react timely. This spectrum of channels, particularly social media, can be quite beneficial for organizations to monitor user perceptions, but also pose some challenges as this feedback have to be actively monitored and trends have to be validated to prevent risk of misinterpretation. 
% Not monitoring this social media feedback could result in snowballing of negative or misinformed user concerns, potentially causing revenue loss. Although we did not find a systematic approach from our participants for managing user feedback, our interviews suggest a secret life of user feedback. A common sequence of activities emerged from our study that we blend as a life cycle of managing user feedback, which consists of activities such as analysis and validation of user feedback for accuracy and understanding. A group of participants also voiced great confidence and efficacy in their organizations' ability to effectively manage user feedback. Specifically, they refer to their ability to understand user feedback and perception and effectively address and respond to these feedback.We report on the best practices that emerged from these participants, which involved taking an active taking a role in employee participation, feedback collection, triangulation, and response to user feedback.  

\end{abstract}

%% Separate the keywords with commas.

% \begin{CCSXML}
% <ccs2012>
%    <concept>
%        <concept_id>10011007</concept_id>
%        <concept_desc>Software and its engineering</concept_desc>
%        <concept_significance>500</concept_significance>
%        </concept>
%  </ccs2012>
% \end{CCSXML}

% \ccsdesc[500]{Software and its engineering}

%https://www.overleaf.com/project/62aaa56ea4263227e92326a1
% \setcopyright{acmcopyright}
\maketitle

\section{Introduction}\label{intro}
% Social media has exploded as a plaza of social discourse for users around the world to engage in conversation about software products, raising awareness about issues or suggesting new features or improvements.
With recent software development trends that encourage rapid iterations and quick feedback loops \cite{nath2018continuous}, understanding user feedback is crucial for software organizations to respond to the evolving needs of customers.
Software organizations face a growing amount of user feedback that come from a variety of different sources \cite{tizard2020voice}. 
While user feedback mainly represent different concerns and interests from various users, these concerns can rapidly gain popularity and become ``\emph{trending}".
Snowballing \cite{arif2016information} can lead to more discussion, attention and sometimes major revenue losses for developing organizations.
Organizations can leverage this feedback to make decisions about new requirements or software improvements.

Analyzing user feedback for eliciting new requirements has long followed an end user-centric approach, where organizations relied on interviews and surveys with stakeholders to elicit key and important requirements \cite{pacheco2018requirements}.
However, the sources through which this feedback is derived from have changed significantly over the years \cite{pagano2013user}.
CrowdRE is the term for research studying processes and tools for collecting and analyzing online user feedback (i.e., the crowd \cite{groen2015towards}).
CrowdRE research has explored additional feedback sources such as app reviews \cite{guzman2014users, pagano2013user, maalej_bug_2015}, product forums \cite{tizard2019can}, Twitter \cite{williams2017mining}, Reddit \cite{iqbal2021mining}, and Facebook \cite{kengphanphanit2020automatic}.

Software organizations still lack a structured process of incorporating user feedback from CrowdRE channels such as product forums or app reviews to their products, unlike in other industries such as retail \cite{ayodeji2019social}, hotel \cite{torres2014stars},  entertainment \cite{bhave2015role}, automotive \cite{Silva_car}, and pharmaceutical \cite{zhan2021social}.
For example, the hotel industry frequently relies on online reviews for insights into guests' general attitudes and feelings, employing this information to improve service quality \cite{torres2014stars}. 
Likewise, social media analytics are frequently used as a key source of feedback for retail companies \cite{ayodeji2019social}.
Organizations can benefit from deeper understanding of user preferences and behavior \cite{hamid_social_nodate}.

We have limited understanding in managing user feedback to improve existing products in the software engineering space. 
Johanssen et al.  \cite{johanssen2019practitioners} reported a 2019 study in this area for companies conducting continuous software engineering from the lens of feedback collection methods, but provides limited insights on how organizations managed the feedback from different sources once collected. Similarly,
Oordt et al. \cite{van2021role} conducted a high-level exploration on the industry practices of using user feedback but provided a limited overview of the process.
% Therefore, more empirical insights supporting the use and management of data from the evermore increasing number of sources of user feedback is needed, along with more structured approaches for analyzing and acting upon feedback.
Therefore, more empirical insights on a detailed and structured set of practices of managing user feedback is needed.

Our research was guided by the question: \textbf{\emph{how do software organizations manage user feedback to help improve existing products?}} 
% We embarked on an exploratory study following a Straussian Grounded Theory approach \cite{strauss1990basics} that aimed to answer this question through interviews with 40 industrial participants from 32 organizations regarding their management of user feedback. 
We report an exploratory study following a Straussian Grounded Theory approach \cite{strauss1990basics} that aimed to answer this question through interviewing 40 industry practitioners from 32 organizations regarding their management of user feedback.
Our study participants came from 9 small, 9 medium, and 14 large organizations in domains ranging from e-commerce, analytics, and gaming, and played diverse organizational roles such as product manager, CEO, data scientist, data analyst, and customer success.

% Several key concepts were identified in our analysis, organizations have to consider user feedback \cite{tizard2022voice, tizard2020voice} to achieve understanding about user perception and concerns and organizations manage the user feedback in a life cycle of steps.
We found that user feedback originates from a variety of different sources, but this variety causes challenges when acting on it. 
% The different ways that organizations manage feedback suggests that there is a secret life of user feedback once organizations collects it from multiple channels, and that social media channels bring opportunities to organizations but they have to be aware of potential pitfalls. 
As a first contribution of our study, we synthesize a number of activities that emerged across organizations for managing these challenges, to create a \emph{life cycle of managing user feedback}. This includes analysis and validation of user feedback for accuracy and relevance to product.
%A sequence of activities emerged from our interviews regarding this management of feedback that we synthesize in a life cycle, which includes analysis and validation for accuracy and relevance to product.
The second contribution of our study is a set of \emph{best practices in managing user feedback} as they emerged from our in-depth analysis of interviews from practitioners that believed their organizations were effective in managing user feedback (i.e. in terms of understanding their user concerns and perceptions and avoiding user dissatisfaction). 
Those are organizational practices and include active employee participation in the user feedback analysis process, and validation of feedback across different channels.

In addition, social media emerged as a prominent source for many organizations. 
Our first two contributions offers insights to help manage social media user feedback amidst the other sources.

To software practitioners, both the life cycle and best practices in managing user feedback represent actionable empirical results that organizations lacking systematic processes for user feedback analysis can leverage to increase their understanding of user perception and behavior for better products thus reducing user attrition. 
For researchers, our study brings awareness of research needed into approaches and supporting tools to support practitioners in detecting and defining user feedback themes and trends, and making automated tools more accessible for organizations.

\section{Background and Related Work}
% This section outlines relevant literature on the current views of user feedback, and highlights the criticality of social media.

Software organizations have long practiced traditional requirements elicitation techniques such as interviews, questionnaires, surveys, etc. to collect feedback from end users \cite{bennaceur2019requirements}.
While traditional methods are needed and effective to a certain extent \cite{van2021role}, they fail to capture the full range of user needs and perspectives \cite{tizard2022voice} as they often involve a one-time or infrequent gathering of requirements.
A growing body of research in the area of CrowdRE has studied ways to analyze user feedback from the crowd, the ``voices of users" \cite{tizard2020voice}, through semi-automated approaches \cite{groen2015towards}. 
Listening to the voices of users helps organizations reap the benefits of feedback \cite{van2021role}

Other industries extensively study the process of understanding user feedback from both traditional and online sources to improve their products and services on a regular basis \cite{ayodeji2019social, torres2014stars, bhave2015role, Silva_car, zhan2021social}. 
For example, studies in the retail pharmacy industry have developed frameworks to analyze unstructured social media data into supportive information to improve operations and service management \cite{zhan2021social}. 
They utilize statistical and sentiment analysis assisted by business analytical techniques to develop actionable information from most-discussed topics and negative comments.
These findings support critical business operations such as customer services, marketing and operations management \cite{zhan2021social}. 
Similarly, the hotel and entertainment industry closely examines online reviews to gain insights on customer feedback to improve their services. 
Therefore, for a software organization, understanding the complex and intricate processes of analyzing user feedback from the newer online sources is worth an organization's time and resources.

In software industry, different sources of user feedback including app reviews \cite{AlAmoudi_app}, forums \cite{wang_subforums}, and vision videos \cite{karras2021potential} have been explored. 
However, understanding users, given the constantly evolving nature of the various sources of feedback is more complex than simply ``analyzing user feedback''; it involves the gathering and analysis of feedback from potentially diverse users with different needs, preferences, and expectations \cite{tizard2020voice}. 
Considering user feedback from a wide variety of sources can bring up complications for interpreting user feedback; one example of this may be considering the culture from which the feedback originated \cite{tizard2021voice}. 
For example, an individual from a more collectivistic culture is more likely to recommend a software product to friends compared to someone from an individualistic culture. 
This phenomenon was especially observable when searching for feedback via social media interactions \cite{tizard2021voice}. 
With the current state of knowledge, software engineering teams do not have the processes to correctly interpret it within the scope of location, culture, and various other demographics. Thus, risk implementing the wrong changes, or overlooking critical feedback.

Growing social media platforms are known as catalysts for new innovation in organizations \cite{ogink2019stimulating}. 
With an increase in social media engagement, users are generating more narratives in their user feedback around software products \cite{li2022narratives}. 
However, managing all the trends that these narratives give rise to is a very complex task. 
% Looking through trends on platforms such as Reddit and TikTok are as close as developers can get to users to gather feedback and honest opinions \cite{price2017eavesdropping}.
Previously, it has been found that organizations with presence on fewer social media platforms also had the smallest proportion of feedback coming directly from users, which contradicts current recommendations to listen directly to the voices of the user's \cite{wiegers1997listening, 9218201}. 
Considering more online channels for feedback helps developers gain a more comprehensive picture of what their end-users need \cite{9218201}. 
Despite the benefits of using social media for deeper insights into user feedback, a previous study revealed that while 9 out of 11 startups look into social media for a competitive edge, larger companies only loosely track their social media presence \cite{10.1145/1984701.1984702}. 
This suggests that many companies are not grasping a comprehensive view of the requirements expressed by their users, and are developing products that may not serve the true needs and wants of their clients.

Previous studies investigating the industry practices employed by organizations have primarily reported on the feedback collection and analysis process \cite{johanssen2019practitioners,van2021role}.
Organizations rely heavily on manual methods for collecting user feedback which is often obtained through explicit channels, such as app reviews. 
Implicit feedback \cite{maalej2009users}, which is unintentional feedback such as user software usage, is often overlooked in the feedback collection process.
Johanssen \textit{et al.} \cite{johanssen2019practitioners} interviewed 24 practitioners from 17 organization and recommend improving user feedback collection through continuous integration, as it has the potential to diversify the requirements organizations work with. 
In addition, Oordt and Guzman \cite{van2021role} through interviews with 18 software practitioners and surveys of 101 practitioners, found that companies are considering newer evolving feedback channels like social media to improve their feedback collection methods.

Despite the efforts of various studies to learn about the utilization of user feedback in the industry \cite{van2021role, johanssen2019practitioners}, they lack in providing a comprehensive and structured set of practices on how user feedback from various sources can be effectively managed and acted on. 
To address this gap, we embarked on this study from a grounded theory lens with the goal of gathering insights from practitioners on uncovering a structured set of activities to manage user feedback. 
% what are the actions they take to manage it internally.

\begin{table*}[h!]
    \centering
      
    \caption{Example Coding of Raw Quotes}
  \vspace{-10pt}
    \small
    \begin{tabular}{p{9cm}p{3.2cm}p{1.2cm}p{2.5cm}} \toprule
       Raw Quote & Code & Category & Concept\\ \midrule
         We do direct feedback from the customer via email, phone call, or chat support. So that can come in through our CSMs, or through our tech support channels. We also have, like an online community that our company built, it's kind of like open source feedback. & SocialMediaUtilization, \par InboundFeedback, \par OutboundFeedback, \par SourceOfUserFeedback &  Collection & Life Cycle to Manage User Feedback  \\  \midrule
         There's a lot of ownership among devs. If you're working on something, it's expected that you own that and see the potential risks and impact of you're doing and make design decisions & ProductManagement, \par  SoftwareDevelopmentProcess & Active \par Employee \par Participation & Best Practices in Managing User Feedback \\ 
         \bottomrule
    \end{tabular}
 
    \vspace{-10pt}
    \label{tab:coding}
\end{table*}

\section{Research Methodology}
% We conducted an exploratory study on the management of user feedback by software organizations.
% A particular focus of our research was exploring the use of user feedback from ``social media'' to help drive product innovation in software organizations. 
To achieve broader understanding of the industry practices for managing user feedback, we adopted \emph{Straussian Grounded Theory} \cite{strauss1990basics} to collect and analyze data from industry practitioners. 
We conducted semi-structured interviews that constantly evolved during the study based on the information emerging from previous interviews.  
We then employed open coding, axial coding, selective coding and constant comparison with existing literature to constantly update our findings and questions throughout the study until reaching saturation \cite{sebastian2019distinguishing}.

\subsection{Participant Recruitment and Selection}
% Our first step involved recruiting prospective interviewees for our study. 
In our strategy we invited from our personal contacts  practitioners working at various positions in software organizations. 
We then extended to networking events and recommendations from other interviewees to increase our pool of interviewees. 
We aimed to talk to practitioners from different sizes of organizations (i.e., a small business has less than 50 workers, a medium business between 50 and 249 workers, and a large businesses has 250 or more workers \cite{oecd}), as the size could play a role in the practices and challenges experienced by an organization. 
We did not solely study large organizations, who likely have more resources than smaller counterparts and are likely using the best practices for managing user feedback. 
As small and medium sized organizations represent the vast majority of businesses \cite{world_bank}, we aimed to uncover discrepancies between the organizations and identify potential areas of improvement.  Our participants came from 9 small, 9 medium and 14 large organizations.
Additionally, we strived to engage practitioners from a wide range of industries to help fill the breadth understanding, and were able to engage with software organizations from industries ranging from e-commerce, analytics, and gaming. 

As our study advanced, we enriched our strategy of engaging diverse organizational roles in our study; our interviews revealed various facets of user feedback, and helped us evolve our recruitment to fill the gaps of our understanding regarding the industry practices. 
While our initial interviews included participants directly associated with product management such as product managers, CTOs, CEOs, and customer success managers, upon reaching saturation from the interviews with the participants related to typical product roles, we learned about the benefit in involving other roles within organizations who are often assumed to be part of development rather than management or product process.
These roles included: data engineer, requirements engineer, and QA engineer.
This resulted in interviewing a wide range of roles, each of whom brought a diverse set of experiences in collecting or managing user feedback as part of the software development life cycle.

Thus, we ended up with a rich set of roles such as those 
(1) involved in user feedback collection (e.g., customer success agent who takes user phone calls) and 
(2) involved in feature development life cycle in some capacity (e.g., product manager who makes final decision to add a new feature).
Table \ref{demographic-table} provides the demographic details of our 40 participants from 32 organizations. 
% The demographic details of our 40 participants from 32 organizations has been provided in the replication package \cite{anonymous_2023_7783840} 
To ensure confidentiality, we have used P\# (P1-P40) to indicate the participants and O\# (O1-O32) to indicate the organizations they belong to.

\subsection{Interview Design}
We conducted semi-structured interviews lasting approximately 30-60 minutes and collected detailed notes and recordings.
% The initial set of interview questions were prepared by the research team following the \textit{general interview guide} method \cite{patton2002qualitative}.
An initial set of 10 interview questions was prepared by the research team, following the general interview guide method \cite{patton2002qualitative}, and based on our understanding of the existing work on user feedback \cite{johanssen2019practitioners,van2021role}. 
We provide our questions in the replication package where the first 10 questions are indicated as base questions \cite{anonymous_2023_7783840}.
We followed the base interview transcript for each interview which includes questions such as \q{What are the sources of user feedback that your company typically uses? Do you feel your organization is effective in managing user feedback?} 
However, as the interviews progressed we adapted our questions depending on the role of the participant to prioritize questions relevant to their role. For example, a product manager has a broad understanding of the function of the product team. Thus, we would ask the question: ``Do you or your product team consider recent trends while identifying features?''. In contrast, a developer may not have a comprehensive knowledge of the tasks carried out by the product team.
As the interviews progressed, and in line with grounded theory practices \cite{strauss1990basics}, our questions evolved with the addition of questions relevant to specific roles or situation so we improvised depending on the participants response. 
% We started each interview with introducing the intention of the research to the participants and asked them to start with discussing their roles and responsibilities at the company. 
% \url{https://doi.org/10.5281/zenodo.7783838}. 
% For each interview, detailed notes and recordings were taken.

\begin{table}[h!]
\centering

\caption{Participants and Their Roles and Organizations}
  \vspace{-10pt}
\setlength\extrarowheight{-10pt}
\small
\begin{tabular}{@{}p{0.5cm}p{1cm}p{3cm}p{1.1cm}p{1.8cm}@{}}
\toprule
\textbf{Org.} & \textbf{P\#} & \textbf{Role} & \textbf{Org. Size} & \textbf{Industry}\\ \midrule
O1 & P1 & Co-Founder and COO & Small & Analytics \\ \midrule
O2 & P2 & Founder and CEO & Small & Analytics \\ \midrule
O3 & P3 & Founder and CEO & Small & Analytics \\ \midrule
O4 & P4 & Product Manager & Small & Educational \\ \midrule
O5 & P5 & Software Engineer & Small & Crypto \\ \midrule
O6 & P6 & Co-Founder and CTO & Small & Food SaaS \\ \midrule
O7 & P7 & Co-Founder and CEO & Small  & Financial \\ \midrule
O8 & P8 & Head of Engineering & Small & CMS \\\midrule
O9 & P9 & Senior Product Manager & Small & Healthcare \\ \midrule
\multirow{3}{*}{O10} & P10 & Cust. Success Manager & \multirow{3}{*}{Medium} & \multirow{3}{*}{E-Commerce} \\
 & P11 & Data Analyst &  \\
 & P12 & Product Manager &  \\ \midrule
 O11 & P13 & Cust. Success Manager & Medium & Content \\ \midrule
 O12 & P14 & Co-Founder and CTO & Medium & Food SaaS \\ \midrule
\multirow{2}{*}{O13} &  P15 & Technical Lead &\multirow{2}{*}{Medium} & \multirow{2}{*}{Analytics}  \\
 &  P16 & Co-Founder and CTO & \\ \midrule
O14 & P17 & QA Engineer & Medium & Content  \\ \midrule
O15 & P18 & Software Engineer & Medium & Healthcare \\ \midrule
O16 & P19 & Data Engineer & Medium & Music \\ \midrule
O17 & P20 & Software Engineer & Medium & Dating \\ \midrule
O18 & P21 & Software Analyst & Medium & Industrial SaaS \\ \midrule
O19 & P22 & Senior Engineer & Large & Automobile \\ \midrule
O20 & P23 & Product Manager & Large & Travel \\ \midrule
\multirow{2}{*}{O21} & P24, P26 & Software Engineer & \multirow{2}{*}{Large} & \multirow{2}{*}{Software} \\ 
 & P25 & Senior Product Manager &  \\  \midrule
 % & P26 & Software Engineer & \\
\multirow{2}{*}{O22} & P27 & Product Manager & \multirow{2}{*}{Large} & \multirow{2}{*}{E-Commerce}  \\
 & P28, P29 & Software Engineer &  \\ \midrule
 % & P29 & Software Engineer &  \\ 
 O23 & P30 & Software Engineer & Large & Security SaaS \\ \midrule
O24 & P31 & Data Engineer & Large & Insurance \\ \midrule
O25 & P32 & Software Engineer & Large & Gaming \\ \midrule
O26 & P33 & Data Scientist & Large & Financial \\ \midrule
O27 & P34 & Software Engineer & Large & MaaS \\ \midrule
 \multirow{2}{*}{O28} &  P35 & Head of Consumer Product &\multirow{2}{*}{Large} & \multirow{2}{*}{Software}  \\
 &  P36 & CTO & \\ \midrule
O29 & P37 & Requirements Engineer & Large & Technology \\ \midrule
O30 & P38 & Technical Project Manager  & Large & Gaming \\ \midrule
O31 & P39 & Data Engineer & Large & Social Media \\ \midrule
O32 & P40 & Senior Software Engineer & Large & Advertising \\ \bottomrule
\end{tabular}

  \vspace{-10pt}
\label{demographic-table}
\end{table}

%Review: One aspect that could be clarified is saturation and how and when this was reached.Although the authors present the sampling logic, a nice addition would be an overview of when saturation was deemed per sub-part of their theory. Given the constantly reflecting and inductive approach, this should be possible to present.

\subsection{Data Analysis}
The analysis process comprised of few different steps.
As we recorded each interview, we transcribed them using an automated transcribing tool that converted audio to text.
The data analysis involved coding process: open, axial and selective along with constant comparison during every step \cite{strauss1990basics}. 
For open coding we broke down the interview transcripts into manageable dialogues and identified initial concepts based on our interpretation of the data.  
Two of the co-authors conducted open coding on the interview transcripts until all the transcripts were codified.
After codifying each transcript, the two co-authors would discuss the definitions of each code and usage of the codes to increase shared understanding. 
In the following steps, we aggregated the codes into broader categories based on different contexts and patterns using axial coding \cite{strauss1990basics}. 
Throughout this process, we continued conducting interviews to gather more evidence on the developed concepts and categories. However, during the last four interviews, we did not uncover any new insights, indicating that theoretical saturation has been reached \cite{strauss1990basics}.
We finally combined the broader categories into one core category through selective coding to reach our final theoretical conceptualization \cite{sebastian2019distinguishing}. 
Table \ref{tab:coding} includes an example of the steps taken during the analysis process.
The two major concepts found from the analysis steps include: life cycle to managing user feedback and best practices organizations engage in for managing user feedback.
A detailed codebook consisting of examples and code to concept generation has been provided in our replication package \cite{anonymous_2023_7783840}.
To ensure participant confidentiality we refrained from including the coded transcripts.

\subsection{Member Checking}
To ensure reliability and checking fit of our findings according to Strauss and Corbin \cite{strauss1990basics}, we conducted member checking with our interviewees. We presented our life cycle of managing user feedback and the best practices for managing user feedback to ten of our interviewees. While they agreed overall with the life cycle steps as they felt they captured the essence of managing user feedback, few of the feedback indicated that some of the steps in their organizations are often merged together (i.e., triangulation and prioritization). 
Several interviewees also pointed out that implementing all four best practices in an organization can be very difficult.
Depending on the organization and with the limitation in the current available tooling, it may not be easy for an organization to collect all available user feedback from all the relevant sources. 
% Several interviewees discussed that our findings would be immensely useful for their organization's management of user feedback and that they would try to implemen.

% In the next few sections, we provide an in-depth analysis of our exploratory study and our interviews with our participants.
% We highlight the key findings about managing user feedback, particularly the social media aspect and what it means to be high a functioning organization for user feedback.
% We also discuss the differences and similarities identified in our study where relevant with previous literature about product management.

\begin{table}[b!]
    \centering
    \caption{User Feedback Sources}
    \vspace{-10pt}
    \small
    \begin{tabular}{p{1.35cm}p{2.5cm}p{3.5cm}}  \toprule
    Category & Source & Organizations That Rely On This Source\\ \midrule
    \multirow{9}{*}{Social Media}   & Reddit & 1-2, 12, 16-17, 19, 21-22, 25, 30 \\ 
    & Twitter  & 1-2, 5, 11, 13, 17, 19, 22, 24-26, 29-30 \\ 
    & YouTube & 1, 23, 30 \\ 
    & Discord & 1, 2, 5 \\
    & LinkedIn & 7, 12, 21, 23, 29 \\
    & TikTok  &  17, 25 \\ 
    & Instagram  & 1, 17, 20, 31 \\ 
    & Facebook & 17, 19, 27, 31 \\
    & Community Forums & 4, 6, 8-12, 18, 21, 22, 30, 31 \\ 
    \midrule 
    \multirow{6}{*}{Conventional}    & Email  &  1-32 \\ 
        & Phone Call  &  3, 4, 6, 7, 9-11, 16, 22, 25, 28, 30 \\ & Video Conferencing & 3, 4, 7, 20 \\ 
        & Support Channels & 1, 4, 5, 9, 10, 11-25, 27, 28, 30-32
        \\ 
        & App Reviews & 17, 22, 25, 26, 27, 28, 30, 31 \\ 
        &  User Workshops &  11, 28, 29, 30 \\ \midrule
        User Usage Metrics & Data from Tools (e.g. Hotjar, Pendo, Google Analytics, etc) & 1-8, 10-11, 13-14, 16-17, 19-23, 25-31\\
        \bottomrule
    \end{tabular}
  
    \vspace{-12pt}
    \label{tab:user-feedback-sources}
\end{table}

\section{Managing User Feedback}
Our participants indicated a rich variety of sources of user feedback that is important to their business, ranging from app reviews, support channels, emails, phone calls, to usage metrics. 
We capture three broad categories of user feedback that emerged from our study in Table \ref{tab:user-feedback-sources} and explain approaches for managing the information from these sources in detail in Section \ref{life_cycle_section}.  
The one emergent category of feedback that manifested throughout our interviews was the feedback stemming from social media. 
Interviewees highlighted to us that social media can have great impact on user perception and misinterpreting can be costly, so organizations that are social media user feedback should continuously monitor and respond to it. 
Our study organizations were quite cognizant of the risks from misunderstanding user feedback, particularly as social media can amplify snowballing of user dissatisfaction, as we describe in Section \ref{Why-User-Feedback-Matters}.

Our interviews showed a lack of a consistent, systematic approach to managing user feedback across the organizations. 
Yet, a number of activities emerged as across the organizations in how they consider user feedback and we describe them in a \emph{life cycle of managing user feedback} and which encompasses the \emph{collection}, \emph{analysis}, and \emph{validation} of user feedback, and its use in the \emph{prioritization of features/bugs} in the software development process (Section \ref{life_cycle_section}). 
The variety and richness of information in the many user feedback channels, although offering unprecedented opportunities to listening to the voice of the users, creates significant challenges for organizations once the user feedback is collected, and there are subtle ways in which it is analyzed, checked for accuracy and validated across multiple sources. 

%However, we identified a series of sequential activities   that we organize in a life cycle to manage user feedback.

%To acquire a rich understanding of how organizations manage user feedback, particularly social media, we interviewed a diverse group of organization sizes and industries.

%While organizations typically collected different variations of feedback sources, they all managed user feedback as they wanted to avoid potential consequences from not considering feedback.

We also performed an in-depth analysis of interviews from practitioners that believed their organizations were effective in managing user feedback (i.e. in terms of understanding their user concerns and perceptions and avoiding user dissatisfaction).
We refer to them as \emph{high performance organizations} henceforth, with the view of distilling actionable insight for the smaller organizations that acknowledged their ongoing struggle and potential shortfalls in their approaches to user feedback. 
These practitioners came from typically larger organizations \org{1, 10, 19, 21, 22, 27, 28, 31}. A number of \emph{best practices in managing user feedback} emerged from our data analysis and we describe them in detail in Section \ref{best_practices}. 
%In contrast, smaller organizations typically acknowledged their potential shortfalls, and where keen in learning more about the best practices. 

\subsection{Why User Feedback Matters} \label{Why-User-Feedback-Matters}
\subsubsection{Misunderstanding User Feedback can have Major Consequences  and Lead to Misguided Decisions and Wasted Resources}
A common theme that emerged throughout our interviews with practitioners was the criticality of \emph{accurately} understanding user feedback.
%Misunderstanding user feedback, especially social media, can lead .
P1 used a movie example as an analogy to illustrate the potential consequences of misinterpreting users. 
\qs{The movie Morbius became a meme... The meme was that no one [saw] this movie. [Discussion about the movie] became very popular. Sony actually extended the theatrical release of the movie, because they saw a trend that the popularity of the movie was increasing. They did not test to determine that the [people] were having fun talking about how no one had [any] intention of seeing the movie.}{P1} 
Consequently Sony wasted money re-releasing the movie to no audiences \cite{mendelson_box_morbius}. 

% This example touches on two crucial aspects regarding social media user feedback. 
This example touches on two crucial aspects regarding user feedback. 
% First, social media user feedback snowballs and can rapidly increase in magnitude \cite{arif2016information}. 
First, user feedback snowballs and can rapidly increase in magnitude \cite{arif2016information}. 
It would be in the interest of an organization to consider the discussion and potentially harness the discussion to its benefit.
% Second, consequences of misunderstanding social media feedback are economic.
Second, consequences of misunderstanding feedback are economic.
If Sony made an accurate assessment of user feedback, they would have realized that no one would spend money on the product, but their attention was just on the pure volume of mentions of the movie.
Hence, our participants emphasized that it is not enough to just consider feedback, they also need to accurately analyze user feedback to understand what is influencing user perception regarding their product. 
This example not only demonstrates ``why user feedback matters'' but also demonstrates how the emerged user feedback category ``social media'' plays a significant role in effective feedback management.

\subsubsection{Social Media Influences User Perception and Behavior}
Our participants emphasized that the increased importance of social media means placing more concern regarding the user perceptions that develop. 
% With the proliferation of social media, user perception can quickly snowball and gain traction. 
\qs{One YouTuber [highlighted a big problem in our game]. In [our game]... for this year, the game team decided to revise how the system works... So there were a lot of negative feedback from the community through public channels}{P38}
The organization had to swiftly address this growing storm of discussion and assuage users about the problem. 

To address the user perception, our participants also discussed its role in shaping user behavior. 
In a previous work that analyzed \emph{narratives} in social media, the authors found user perception having a major influence on user behavior. For example, where software products perceived as suffering from a lack of privacy in turn experienced users flocking to competing platforms \cite{li2022narratives}. As P35 put it, \q{user behavior [also] spurs research and insights and feedback to a large degree, you know, because you can take that and mold it and shape it and put it back into a product road map.}
% In other words, when a user struggles to even understand the main features of a software, it is likely that users will complain or refuse to use the product.

User complaints are problematic because the organization will have to address them to diffuse negative feedback.
P31 explained the measures that their organization took to increase the positivity from user perception on social media.
\q{[We] used to analyze the perception. So most of the times the feedback is negative on Twitter. [Redacted] made efforts to improvise that image in public by analyzing the tweets and then taking action.}
Upon releasing new features that addressed major concerns, the number of negative feedback decreased. 
The reduction in negativity in user perception was a result of carefully curated efforts to deliver features that were well received by users.

\begin{figure}[b!]
  \centering
  \vspace{-15pt}
  \includegraphics[width=1\linewidth]{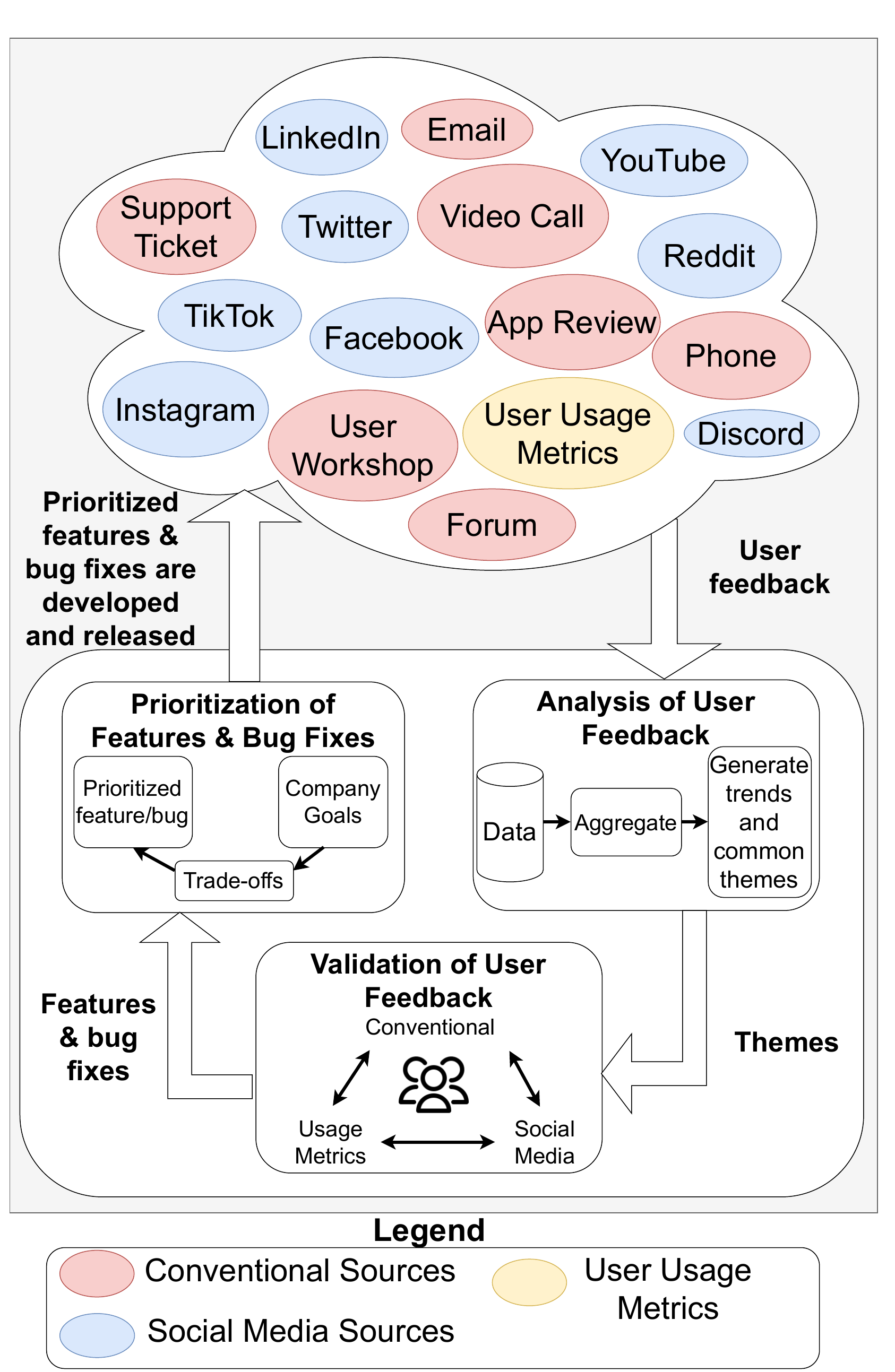}
  \caption{Life Cycle to Manage User Feedback}

  \label{voice-of-users-data-flow}
    \vspace{-10pt}
\end{figure}

\subsection{The Life Cycle of Managing User Feedback} \label{life_cycle_section}
The consequences of not acting or acting wrongly on user feedback and perception are both significant and best avoided. From our data we synthesize a life cycle to manage user feedback (Figure \ref{voice-of-users-data-flow}) for software organizations based on our practitioner interviews and previous literature.

\subsubsection{Collection of User Feedback:}\label{collection-lifecycle}
The three broad categories that emerged from analyzing the transcripts of our interviews, consisted of conventional, social media, and user usage metrics. 
We found that depending on the feedback category, the organizations manage them quite differently.
This difference largely originates from the provenance of each user feedback as methods for collection and the resulting data type is different for each feedback category. 

For conventional user feedback, a user is providing direct user feedback to a particular software organization in the form of a phone call, video conferencing, or user workshop.
Unlike social media sources where it is unclear whether a user actually uses a product, an organization receives feedback from someone who is likely using the product with conventional user feedback.
The conventional feedback will often be in the form of textual transcripts of user input from sources like support tickets and emails. These sources will also often include photos or screenshots of the software, which organizations will collect into issue tracker systems.
Sources from conventional sources include the most popular source, email, as every organization listens to email feedback.
Moreover, support channel communications and phone calls are also frequently leveraged by many organizations.   

Similar to conventional sources, a vast majority of organizations \org{1-2, 5, 11-27, 29-31} emphasized the importance of collecting user feedback from social media sources.  
% Notwithstanding organizations that do not receive feedback from social media, the vast majority of organizations \org{1-2, 5, 11-27, 29-31} discussed the importance of collecting social media user feedback to understand issues and needs of the product.
% \qs{We have a team that just looks at social media [feedback], and they report on the biggest trends coming from social media.}{P27}
We found that organizations are increasingly considering social media feedback, especially from newer types of social media such as TikTok \org{17, 25}, Discord \org{1, 2, 5}, and Instagram \org{1, 17, 20, 31}.  
While previous studies on user feedback have tried to provide greater understanding on the types of feedback, they did not specify the details on how organizations actually collect and utilize the social media feedback \cite{van2021role}.

We found that large organizations often paid close attention over social media, which is reasonable as they typically generate a lot of media attention.
\qs{You can imagine this company is in the news all the time. It's definitely a part of our weekly life of what we look at.}{P27}
In contrast, many of our smaller organizations do not consider social media as they have not reached the scale to experience a large magnitude of feedback \org{3, 4, 6, 7, 9}.
Our finding contradicts earlier research on the proliferation of user feedback in organizations, which indicated that small organizations consider social media, where as large ones do not \cite{10.1145/1984701.1984702}.

However, the smaller organizations that do consider social media in our study have extensive experience and shared some of the pitfalls and mitigation strategies.
For example, both O1 and O2 heavily rely on a combination of social media to build understanding of the user base and conceptualize new features.
\qs{Before we jump in and spend \$40,000 plus on developing that functionality... let's get a sample of the social media posts, let's get a sample of the events in question.}{P1}
Acknowledging that there could be bias on social media, P2 shares a practice they use to reduce bias in their data collection.
\qs{Generally I don't ask question [to the community] under my name... I have some fake name with a male and a female account. [One] needs to stop being biased. Some guys want to seduce a girl and some guys are gonna be aggressive with us. So I'm basically trying to have neutral data points, non biased.}{P2}
In other words, successfully managing user feedback, especially from social media, starts with collecting non-biased user feedback from users.

% In addition to social media, organizations also utilized numerous other sources of feedback such as user workshops, app reviews, and community forums, which is in line with previous studies in this area \cite{van2021role}.

For social media feedback, a user typically provides feedback in the form of video (e.g., YouTube or TikTok) or textual (e.g., Twitter or Facebook).
Social media feedback is also often collated in the form of text or image, but social media requires a lot more active monitoring and this process for monitoring is very much ad-hoc.
Sometimes, social media feedback is more amplified into a trend due to the fact that \qs{if someone starts talking about a particular feature, more people will start talking about that feature, that doesn't necessarily mean that is a valid signal, they're talking about it because of conversation is occurring}{P1}. 
Recall, the example with the movie Morbius, people mentioned the movie on social media, but the movie studio had no way to know if these people were actual moviegoers or not.

Finally, we categorize data collected by tools such as Hotjar or Google Analytics in the category user usage metrics.
Tools like Hotjar provide comprehensive insights on the way that users use a product ranging from button clicks to mouse hovers.
These tools are sophisticated enough so that anyone at the software organization could replay a user usage session and visualize every action. 
\qs{Everything is tracked, every click is [tracked], every action is tracked. That user feedback is how users interact with the [product].}{P39}
With user usage metrics, the source of feedback more accurately represents user actions and does not depend on what users are telling the organization.

As shown by Table \ref{tab:user-feedback-sources}, different sources are used by different organizations. 
The sources used depends on several factors including ease of access, cost, availability of tooling and data.
For example, small organizations may not have reached critical mass to receive significant amounts of user feedback.
Therefore, it is important for organizations to weigh their current needs and determine the critical sources. 
% Regardless of the category, important feedback eventually gets added to issue tracking systems, \qs{We have different Jira [tickets that are created] when someone [on] feedback collection team collect some requirements or feedback.}{P19}

% We found that organizations treated conventional and social media types differently. 
% With conventional feedback, an organization would be talking or observing a verified user, whereas this may not be the case for social media users. 
% Another difference manifests in the availability of data.
% Social media, particularly in small companies, may not be available. 

%Tt is important to note that the choice of data collection tools and methods may vary depending on the specific goals, context, and constraints of each organization, 

\subsubsection{Analysis of User Feedback:}
Our participants indicated the next step after collection is analyzing the trove of feedback.
Key to the process is listening to the user feedback channels and group similar patterns.
\qs{[We] went through all that feedback and grouped the similar themes about what were the common complaints and what were the common dream features that everybody wanted.}{P4}
\qs{The number of times the question was raised was also important, because that means many [users] are having this confusion.}{P5}

% \begin{wrapfigure}{l}{0.12\textwidth}
%   \centering
%     \vspace{-10pt}
%   \includegraphics[width=0.16\textwidth]{figures/analysis.pdf}
%     \vspace{-20pt}
%   % \caption{User Feedback Triangulation}
%   \label{validation}
% \end{wrapfigure}

%indicating that the analysis process is somewhat same but differs in reliability
For both conventional and social media user feedback, organizations tend to consider feedback as a ``theme'' when it gains traction. 
\qs{You got a trending amount of, let's say support tickets complaining about a specific issue}{P14}
A problem emerges as a theme when there is a growing or significant amount of feedback.
However, organizations may view conventional sources as more credible, as social media feedback may be noisier.
In the case of user usage metrics, organizations look for patterns through graphs that indicate a trend. 
\qs{Pendo [a kind of product analytics and feedback collection tool,] is kind of cool. Because it tracks every click, you can go into it on a user level and see exactly where someone clicked, at what time, and they generate these fantastic graphs for you.}{P10}

%What happens to the generated themes. How themes are sent to next phase.
Once the common ``themes'' are generated, they are sent to the next phase through different methods such as Jira tickets.
\qs{Once we have kind of enough information... who is it's going to affect? Is it affecting every user, how big of a deal is this? The product team creates a [Jira] ticket... and then we work on it and scope it out and do our research. [And] then it is forwarded from there.}{P4}

%How social media is a good addition to the data
The perks of using social media user feedback includes rapid insights on bugs that occur in the product.
\qs{During the testing process, we missed [including the menu items]. We woke up one day... All these messages on social media being, ``what the heck, I just updated and none of the menu items [exist] anymore." }{P13}
Monitoring social media, if done effectively, can be powerful for quick feedback on bugs in the software.
We see the importance of pure volume of feedback manifesting as a key factor of identifying themes.
\qs{People will post snapshots of a Reddit post with a whole bunch of votes. And [users are]  mad about the [feature], or on the new UI. And then a month later, there's a JIRA ticket and someone's working on it.}{P22}

A challenge with monitoring social media user feedback is the presence of noise, particularly in social media where spontaneous comments can quickly gain momentum. 
P38 emphasizes that it is essential to balance the signal-to-noise ratio and consider which groups of users are over-represented or under-represented in this type of feedback. 
After all, some groups may be more likely to post on social media than others, which can skew the data \cite{tufekci2014big}.  
It is important to ensure that the feedback received represents a diverse range of users and is not biased towards any particular group.
It is crucial to listen to all feedback, but it is equally important to recognize that not all feedback is equal, as feedback can suffer from poor quality \cite{van2021role}. 
% Oordt and Guzman \cite{van2021role}, reported practitioners echoing similar concerns about ``feedback quality''. 

The process of analyzing social media content is predominantly carried out manually by individuals or teams who actively monitor social media platforms during their regular activities. 
Many of the successful companies have teams that conduct this analysis.
However, smaller companies struggle with the process and prefer to have a tool that would support their team. 
Moreover, the organizations heavily rely on manually looking through incoming user feedback to uncover underlying patterns and trends not just from social media, but also for conventional user feedback as well.
% This is in line with literature \cite{van2021role}, where they reported that industry practitioners tend to incline more towards manual analysis methods.
Existing literature \cite{van2021role} also reported the reliance on manual method and the limited usage of automated feedback analysis among practitioners, despite a significant number of tools developed \cite{carreno2013analysis, guzman2014users, iacob2013retrieving, maalej_bug_2015, oehri2020same, panichella2015can, gao2022listening}. 

Our participants reported similar limitations, but also specified that they do not use existing tools as they are too hard to use, expensive, or still not publicly available.  
For example, O4 attempted to use an existing tool to analyze their community forum in a better way. 
However, the tool was not as effective as expected and had an expensive yearly subscription cost.
Our findings indicate there is high emphasis on identifying reliable ``themes", albeit through manual analysis, so there is strong demand for tools that can enable reliable and cost effective feedback analysis.

\subsubsection{Validation of User Feedback:}\label{sec: validation}
Validating the emerging ``themes" that arise from the analysis phase into a ``list of actionable feature and bug fixes'' for an organization to work on is the next important step.     
% \begin{wrapfigure}{l}{0.12\textwidth}
%   \centering
%     \vspace{-10pt}
%   \includegraphics[width=0.16\textwidth]{figures/validation.pdf}
%     \vspace{-20pt}
%   % \caption{User Feedback Triangulation}
%   \label{validation}
% \end{wrapfigure}
The validation process involves triangulation between the 3 categories of user feedback to reduce misinterpretations.
\qs{In theory, if a [social media] trend is significant, you should be able to see the impact of that trend in your own user observation data as well. And if you can't, that's either you're not doing very good user observation data, or the trend is not impacting you yet.}{P1}
One can compare the themes from conventional and social media sources with user usage metrics and vice versa. 
Another possible strategy is starting with social media to identify themes and then validating the themes using conventional user feedback such as video or phone calls with users to acquire direct feedback. 
\qs{Social media is a good place to get started. You find a trend. Next step is to collect a statistically relevant, smaller sample of genuine accounts that you care about.}{P1}
However, similar to the analysis activity, triangulation has primarily relied on manual means in these organizations.
After triangulation, an organization would have collated a list of feature and bug concepts that were validated and represent genuine issues.

By tracking user engagement with different tooling such as Hotjar or Google Analytics, one can see which features are being used effectively and which ones are being ignored. 
Social media, in particular, come with a lot of noise as discussed previously, but triangulating the themes from social media with user metrics can help understand the user behavior better.
\qs{Feedback is a gift... The metrics don't lie. Metrics are metrics, they have no face... Sometimes it's about how a new feature or change you made, sometimes it's about a bug, you always look for what's the cause and effect of a change in a metric. So you have to kind of look at it from that aspect and realize that you can't always get [feedback] from direct comments from users.}{P36}
A study suggests that industry practitioners rarely applied systematic validation of user feedback \cite{johanssen2019practitioners}. 
While existing literature adds various feedback collection, analysis, and prioritization processes in depth \cite{van2021role, seyff2017end, firesmith2004prioritizing, gartner2012method, johanssen2019practitioners}, there is a dearth of guidance on how to determine the validity of the emerging ``themes".

One challenge with triangulation is the pure volume of data that is produced from the user feedback channels, particularly from social media.
This high volume of data is difficult to validate as this process is often ad-hoc. 
Moreover, the volume of data may even require organizations to dedicate new roles and teams to manage this information.
Therefore, the effort in establishing tools and processes to support this triangulation would seem worthwhile for organizations as it increases the likelihood of identifying which features and bugs are relevant and impactful for users before investing valuable engineering and product hours.

\subsubsection{Prioritization of Features \& Bug Fixes:}
Our interviewees highlighted that prioritization is a crucial step in the life cycle to manage user feedback, which involves making trade-offs between different bugs and features. 
A critical factor to prioritization is whether the ``list of actionable feature and bug fixes'' received from the validation stage match a company's vision and goals. 
Best described by P27, \q{if we get feedback, that doesn't align to anything we're doing... we're not going to redesign it, we're not going to revamp it.}

% \begin{wrapfigure}{l}{0.12\textwidth}
%   \centering
%     \vspace{-10pt}
%   \includegraphics[width=0.16\textwidth]{figures/prioratization.pdf}
%   % \caption{User Feedback Triangulation}
%   \vspace{-20pt}
%   \label{validation}
% \end{wrapfigure}

The underlying driving force for major features is often the revenue that a feature may bring to the organization.
\qs{If there is a feature that one customer wants and [has] a significant impact on revenue, we're going to implement [it]... If they ask to put a pink elephant on the homepage, we're not gonna do that. But [we will] if they [want] a feature that's gonna benefit [many] people.}{P2} 
Organizations place emphasis on activities that generate revenue or prevent financial losses, while also factoring in practical considerations.
This is in accordance with literature \cite{berander2005requirements, sivzattian2001linking} where cost and risk have been indicated as major aspects of requirement prioritization. 

The key to prioritization is balancing trade-offs so that there is mutual benefit for the users and organization. 
P6 astutely observes, ``\textit{if they [users] are giving us [feedback], it's likely that [something] would help them. [They] help us because they are going to use our system.}"
\qs{People's perception of the company as a whole}{P28} is also an important factor. 
If the feedback is related to something that is damaging the company's reputation, it will be given priority even if it doesn't have a direct financial impact. 
After all, a company's reputation is invaluable and can influence user behavior to adopt or drop a software product \cite{li2022narratives}.

Our participants also described that they most often prioritized bug fixes over features. 
Bugs are almost always prioritized over feature requests as bug indicate a blocker to current revenue, a top priority issue, where as new features is usually meant to obtain future revenue.
\qs{New features are typically lower priority than bug fixes. So usually bug fixes are prioritized top of the list, because it's something that the users are paying for or having issues with as compared to something that just doesn't exist.}{P17}

The often ad-hoc process of prioritizing can result in the wrong decisions, which is particularly troublesome for smaller organizations with limited resources. 
\qs{Sometimes we end up choosing such a [feedback] to implement, which probably just takes a lot of time. [Alternative] task which [ended up] more important gets delayed. So we do make mistakes, but we try to avoid that by discussing and analyzing.}{P6}
Multiple approaches have been suggested to aid in determining the priority of requirements \cite{achimugu2014systematic, firesmith2004prioritizing}. 
A number of studies have focused on automating the end user feedback prioritization process \cite{licorish2017attributes, kifetew2021automating, gartner2012method, malgaonkar2022prioritizing, gao2022listening}, but as reported by our interviewees, industry practitioners rarely use the tools.  

% Prioritizing user feedback is a necessary step that requires balancing trade-offs for features and bugs.
% By striking this balance, an organization may create software that not only meet user needs but also align with their larger vision and goals.

\begin{table}[!b]
    \centering
    \small
    \caption{Best Practices for Managing User Feedback}
    \begin{tabular}{p{2.0cm}p{5.7cm}}
        \toprule
       \textbf{Practices}  &  \textbf{Criteria} \\ \midrule
        Active Employee Participation  & Every employee is actively contributing to the life cycle to manage user feedback (i.e., collection, analysis, validation, and/or prioritization) \\ \midrule 
        Active Feedback Collection &  
        Proactive data collection from all the feedback sources related to the product.  (i.e., Sources equal \emph{n} where \emph{n} is number of sources containing an organization's feedback and $\emph{n}\geq$2 )
           Sources must come from at least 2 out of 3 feedback categories \\ \midrule
       Active \par Triangulation\strut   & Organization actively leverages 2 or 3 categories (i.e., social media, conventional, and/or user usage metrics) to validate the ``collated list of bugs and features'' \\ \midrule
       Active Response  & Prompt response to the feedback. Response could be in terms of marketing, communication and/or bug fix or feature roll-out.  \\ \bottomrule
    \end{tabular}

    \label{tab:high_performer}
    \vspace{-5pt}
\end{table}

\subsection{Best Practices in Managing User Feedback} \label{best_practices}
Participants from \emph{high performance organizations} shared with us one or more organizational practices in managing user feedback and which they described made them feel confident in achieving a good understanding of users, and in their ability to address their most pressing needs. 
We refer to these \emph{best practices in managing user feedback} as the \textbf{4A}s: 1) Active employee participation, 2) Active feedback collection, 3) Active triangulation, 4) and Active response time.  
We outline them in Table \ref{tab:high_performer}, together with satisfying criteria. 
Many participants (11 out of 32 organizations) indicated the need for these practices, \qs{You know in startups, because the real struggle is we don't know the best practices out there.}{P6}

\subsubsection{Active Employee Participation:} \label{sec: active employee participation}
Organizations described from their experience that high performance in managing user feedback depends on an environment that encourages employees to thrive in user feedback management.
When employees in different management and development roles actively participate in the feedback collection, analysis, validation, and prioritization, the team can effectively incorporate all three categories of user feedback. 

We found that large organizations \org{19, 21, 22, 27, 28, 31} often heavily encourage their employees to take the initiative, particularly for identifying issues in the software products.
\qs{There's a lot of ownership among devs. If you're working on something, it's expected that you own that and see the potential risks and impact of you're doing and make design decisions.}{P22}
Employees who are invested in the product may pick out themes in user complaints on social media while browsing,
\qs{High majority of [our company] employees are also very interested in the product. It's very common that the employees will frequently visit [our company's] subreddits. Bug reports and feedback comes from internal employees being on [our] social media.}{P22}
Employees will frequently post screenshots from social media into internal communication channels regarding popular concerns. \qs{[We see] screenshots of a Facebook post saying ``what does this alert mean?" Then someone's on it, trying to make the customer facing UI more understandable or adding something. This initiative is definitely led by employees that have other jobs that just happen to be also invested in these kinds of things.}{P22}

Participants  from \org{19, 21, 22, 27, 28, 31}  suggested that organizations may benefit when they empower employees to speak up and advocate for certain features that they think are necessary.
\qs{The philosophy at [our company] is there's a significant amount of personal ownership over everything you do... It's a little more effective to bring [decision making] closer to the source.}{P28}
The participants explained that managers often relied on the tacit knowledge and collective wisdom from the team to share expertise about the software to make decisions.
\qs{I [consider the] sources of feedback and then using my contextual knowledge about how important I think this is, we can choose to fix it before it becomes a problem... I [get my managers to] fight for this in the next meeting.}{P28}

Unlike larger organizations where active employee participation was more commonly reported, only one small and one medium sized organization \org{1, 10} expressed utilizing this practice. 
% to identify issues in the products.
The others instead relied on more top down approaches instead of taking advantage of the tacit knowledge and collective wisdom. 
\qs{The product owner would decide the prioritization of what the backlog looks like. [In] their sprint planning meetings, they would pull it into work to be done. I do not let the developers decide the priority.}{P14}.
\qs{I think [once] the product team [sic] has their vision as to what exactly it is that they need, then they'll start pulling the technical team because there's no point pulling us in any earlier [until] they know what they actually want.}{P15}
However, organizations \org{19, 21, 22, 27, 28, 31} that fostered individual empowerment in addressing user feedback indicated that they were able to reap benefits of achieving company goals and revenues.

\subsubsection{Active Feedback Collection:}
As described in the Section \ref{life_cycle_section}, all of our organizations actively collect feedback.
However, for higher performance, the organizations suggest active feedback collection from multiple user feedback categories, which 29 out of 32 organizations practice. 
To effectively gather and incorporate feedback from users, these organizations identify and utilize at minimum two sources that are representative of at least two of the feedback categories (social media, conventional feedback, and user usage metrics). 
To excel in active feedback collection, a participant would also regularly review and update its feedback channels pool to ensure it reflects the changing needs of the organization and its customers.

In addition, as highlighted by many participants, they benefited from having dedicated roles or teams to collate all the various user feedback sources.
Despite data scientists and data analysts roles becoming ubiquitous in these software organizations, the volume of user feedback is quickly becoming vast.
\qs{You have logs for everything. There's almost too much data. I think that one of the biggest issues is who's interpreting the data who's making the dashboard.}{P22}
These organizations typically had to allocate resources to not only collecting user feedback, but also interpreting the data. 
All the large organizations with resources have dedicated teams and roles that serve as the first line to collect the different sources. 

Sometimes these dedicated teams exist in the form of marketing or public relations teams. 
\qs{I think there are a couple teams that are dedicated to look into those. ... like what are the users talking about? What do they like? What do they don't like? ... I'm always getting those emails saying users really love this feature.}{P26}
Additionally, organizations could also benefit from roles that are cross cutting in nature and who understand the various sources of user feedback such as a customer support agent who has knowledge about user usage metrics in addition to conventional user feedback sources. 

As a whole, the participants reinforced the importance of resource allocation within an organization in the form of roles for successfully identifying the related sources. 
By utilizing multiple sources of feedback our high performance participants gained valuable insights and understood the needs and preferences of users.

\subsubsection{Active Triangulation:} \label{sec: active triangulation}
Triangulation emerged as a critical activity in the life cycle of managing user feedback \org{1, 4, 7, 10, 11, 13, 19, 20, 21, 22, 25, 27, 28, 29, 31}.
Considering the high number of participants that practice active triangulation of user feedback, our findings suggest that active triangulation is a best practice for validating \emph{themes} in user feedback.

In particular, we found 15 out of 32 organizations that conducted active triangulation, validated its \emph{themes} from a feedback source with at least one other source from a different category of user feedback.
By following steps described in section \ref{sec: validation}, organizations can verify whether a specific user feedback has validity or not. 
For example, if an organization notices that users are complaining about the UI of the landing page on Discord and determines that it is an emerging theme, the organization could triangulate this theme with user usage metrics to gauge whether or not users are spending less time on the landing page or bouncing from the landing page. 
Triangulating the two categories of feedback in this example, social media source and user usage metrics, would significantly improve the validity of the theme.

Additionally, organizations actively practiced validation before making any decisions about the feedback.
\qs{If someone says, this is a problem, we have a lot of data sources that we can then query to verify that is a problem. And like, hopefully, if someone's DMing, you put in some amount of work to justify this as a priority.}{P22}
The process of triangulation would also require coordination amongst different roles and teams in the organizations. 
Data teams would communicate findings from user usage metrics with findings found from product, customer support, and customer success teams who are more closely in touch with conventional and social media sources. 
This aspect relates to the aforementioned factor in section \ref{sec: active employee participation} ``active employee participation" where \emph{every} employee is encouraged to be cognizant of the feedback and product.
This may open new collaborations and communication channels with cross-functional teams as different teams share responsibility in validating user concerns. 
\qs{I have to bug the [product people] often [for] validating because there are customers that are louder than others, you have to factor in that they could just be one very squeaky wheel. If one customer complains about it, [other] customers behind the screen [may have same issue].}{P12}
Participants discuss facilitating a culture for coordination and collaboration to help support validation.

\subsubsection{Active Response:} \label{active_response}
Our participants who expressed confidence in managing user feedback also suggested active response \org{1, 2, 7, 10, 11, 12, 14, 16, 17, 19, 20, 21, 22, 25, 27, 28, 30, 31} to user feedback as a best practice. 
Active response implies responding to a feedback as quickly as possible.
The response can be either a message or marketing post to the end users addressing the issues.
It could also correspond to fast development and release of the feature.
In both cases, the time of response varies depending on the complexity of the user feedback. 

Participants explained it is insufficient to just consider important user feedback. They need to also respond to user feedback, particularly critical feedback, as soon as possible.
We know from literature that user feedback if left unresolved, can lead to cascading consequences that impact an organization's reputation and even bottom line \cite{li2022narratives}. 
This sometimes means that an organization shifts priority due to a sudden emerging user concern.
For example, P1 highlights a notable example where responding to important user feedback and shifting priorities in a timely manner is paramount. 
\qs{The [product] security updates that occurred following [major] criticism [on social media].
[It] very quickly shipped some very good community safety and encryption changes. It came down to a matter of prioritization... Priority changed, and its message to customers changed because it's quite a bit easier to change priority and messaging.}{P1}

This approach shares similarities to the principles of DevOps, which principles encourage an ``organization's ability to deliver applications and services at high velocity." \cite{aws_what_2023}
In the case of DevOps, organizations are measured by their lead time for a change among other metrics to gauge an organization's ability to release code in small iterative batches. 
However, for active response, it does not necessarily always mean that an organization must ``release" new code right away.
An organization's response to user feedback could be in the form of a new version of software, but it could take form in its messaging or marketing.
As explained by P13, customers value the prompt response even if the product roll out takes time, 
\qs{[Following up] on customers, they really appreciate that... I think, sometimes they paint these stories in their brain that [customer support is] going to take days to get back to me, there's going to be so much back and forth. I think something our customers are always a bit surprised by is: `you actually follow up with with me.'}{P13}

The perceived benefit of applying active response is that users are more satisfied when an organization displays care for user concerns.
Previous study on user-developer interaction on app reviews reports that users tend to increase their ratings after they receive a response from the product page \cite{srisopha2021should}. This is inline with our findings that user perception is positively influenced from proactive engagement.

%Review for active response: App stores offer developers the opportunity to engage with users by responding to their comments, which can help build a trusting relationship. However, it’s unclear how this approach compares to the results (section 4.3.4) of the interview study. I personally think a bit more discussion is needed on how it relates to the topic of engaging with app users through comment responses. 

\section{Threats to Validity}
We use the total quality framework of Roller \cite{roller_applied_2015} and its elements of credibility, analyzability, transparency, and usefulness, to assess the threats to validity and mitigation approaches in our study. 

For \emph{credibility}, our study may suffer from sampling bias as we could only talk to participants who agreed to interviews with us.
However, our interviews indicated that all of our participants manage user feedback in some capacity, and a non-trivial number of participants believed that they were effective in managing user feedback.
We also tried limiting bias by informing the interviewees at the beginning of interviews that participants would be anonymized and the study would not cause risk to them. For \emph{analyzability}, we utilized tooling to assist with transcribing the audio into text and also manually verified the transcripts against the audio.
Two co-authors followed the steps of grounded theory and conducted open, axial, and selective coding to analyze the transcripts.  For \emph{transparency}, we attempted to provide rich descriptions and quotes where possible, and make available a replication package containing our base interview questions and codebook. Due to confidentiality agreements, we cannot release the interview transcripts. 

For \emph{usefulness}, our study is intended to shed more insights on how software organizations manage user feedback. We provide more empirical results regarding the life cycle of activities and best practices that they utilize. 
We conducted member checking with 10 of the participants and presented all the findings that emerged from the study in checking that our research findings resonate with our study participants and their organizational practices. 
We do not expect our results to hold true for all software organizations, though we would expect organizations of similar demographics to those in our studies to share similarities. 
% Additional studies like ours could be useful in providing more empirical grounding for the activities and practices leveraged by organizations. 

\section{Discussion and Conclusion}

Prior work \cite{van2021role, johanssen2019practitioners} reported a high level overview of user feedback collection and analysis practices.
However, we lacked a detailed and structured set of practices for organizations to follow. 
Our study aimed to address this gap, via a grounded theory study with a diverse group of 40 practitioners from 32 organizations of various sizes and in several domains of software development.
Our results yielded a life cycle to manage user feedback and best practices in managing user feedback.

In addition to the life cycle and best practices, we found differences with prior works on how organizations collect feedback. 
% Those often focused on two types: explicit and implicit feedback, whereas our results suggest that there are three main categories of feedback: social media, conventional, and user usage metrics. 
Maalej et al.  \cite{maalej2009users} categorized user feedback into \textit{explicit} and \textit{implicit} to account for the broad spectrum of feedback channels.
Our study refines this categorization into three main categories of feedback: social media, conventional, and user usage metrics, to better reflect the difference in the methods of managing each category.

Previously, the use of social media user feedback has been championed in other domains of research. 
For example, in the airline industry customers who leave complaints and find resolution on social media are more satisfied \cite{socialmediacomplaints}.
% This was also found in our study as exemplified in Section \ref{active_response} where a participant explained how customers really appreciate prompt responses.
Our interviewees also highlighted that social media is increasingly an important source of feedback, but they also warned 
that social media feedback may not represent verified users and noise can be difficult to filter out.
Regardless of these limitations, social media may be a powerful source that can benefit an organization, if social media sources pertain to the organization.

One of the main contributions of our work is the \emph{life cycle of managing user feedback}.
Other domains such as pharmaceuticals \cite{zhan2021social}, retail \cite{ayodeji2019social}, entertainment \cite{bhave2015role}, and hotels \cite{torres2014stars} have also reported on utilizing user feedback to assist in understanding user concerns and behavior. 
The pharmaceutical industry \cite{zhan2021social} uses social media to identify the most discussed topics from customers and identify potential areas for improvement based on negative complaints. 
Some studies from other domains provide suggestions of how to collect data from online sources and make actionable insights \cite{zhan2021social, xiang2017comparative}, but these studies are more focused on the analysis activity as opposed covering the entire life cycle.

% While offering new opportunities to companies, user feedback also requires validation for accuracy against data collected through other channels. 
% While offering new opportunities to companies, social media feedback also requires validation for accuracy against data collected through other channels. 

Additionally, organizations shared with us \emph{best practices for managing user feedback}. 
Unlike prior literature \cite{van2021role, johanssen2019practitioners} where they describe some of the activities that organizations employ for feedback collection, we provide guidelines for four practices that may help in achieving a comprehensive understanding of users.

These practices were more common to large organizations in the study, suggesting a change in the state-of-practice from a decade ago that large companies only loosely tracked their social media presence \cite{10.1145/1984701.1984702}. 
Both these two contributions represent empirically developed actionable insights that organizations lacking systematic processes for user feedback management can leverage to increase their understanding of user perception and behavior for better products and to reduce user attrition. 
Next, we further discuss these implications for both the practice and research of software engineering. 

\subsection{Implications for Practitioners}
\textbf{Social Media User Feedback:}
The unprecedented, unique opportunity afforded by social media is that organizations not only can catch user dissatisfaction right away, but it also supports organizations to address and respond to users in a open manner.
For example, a recent Reddit post sparked a discussion on tenfold increase in prices for the Pro tier version of Google Colab without informing the users \cite{reddit_2023}. 
This conversation was promptly noticed by an employee from product at Google Colab during his regular online browsing as he mentions in response to the post \q {I mostly lurk [on Reddit]}. 
The employee immediately addressed this issue acknowledging that the price increase was a bug in their update, quickly rolled out the bug fix, and issued refunds to anyone who was charged incorrectly. 
The fast response was appreciated by users who admitted that it prevented a significant number of subscription cancellations.
This example illustrates the power of leveraging social media user feedback and addressing and responding to the issue in an open and transparent manner.

\textbf{Best Practices for Managing User Feedback:}
The example demonstrates characteristics of active employee participation that emerged out of our study.
Our findings indicated that all our participants manage user feedback in some capacity, but few organizations follow all the 4As of best practices.
These best practices emerged from our participants that consider their companies high performing in managing user feedback, which suggests that other organizations may benefit from applying these practices.   

% Firstly, for organizations that currently do not manage user feedback in a systematic method, they should consider implementing the life cycle to manage user feedback: 1) collection, 2) analysis, 3) validation, and 4) prioritization.
% And secondly, organizations should employ a company culture where the 4As (factors) of becoming a high performing organization can be enforced.
% Organizations need to be cognizant about these two aspects as they are interconnected. 
% Notably, ``Active Employee Participation'' plays a crucial role when it comes to excelling at each step of the life cycle of managing user feedback.

%Review: Implications for researchers could also be expanded to explicitly consider the different parts of the RE and inform and reflect on directions.

\subsection{Implications for Researchers}
\textbf{Impact of Organization Size and Maturity:} 
Our findings indicated the best practices to manage user feedback. 
Most of our larger organizations employed these practices, whereas smaller organizations were often seeking the best practices. 
Further empirical research should investigate the impact of organization size and maturity on the use of these best practices and whether or how they can be refined for most effective implementations in organizations. 

\textbf{Life Cycle Activities:}
% We list 4 activities in our life cycle to manage user feedback.
Although some prior works exist pertaining to the collection and analysis of user feedback from requirements engineering \cite{maalej_bug_2015, gao2022listening}, future work should explore  utilizing the amalgamation of various sources. As emerged from our study, social media sources are also increasingly important, albeit the presence of noise and sheer volume of data. In addition feedback validation and prioritization should also be further studied in the context of the evolving nature of feedback sources. 

\textbf{More Tools are Needed:} Practitioners in our study reported that they heavily relied on manual approaches for both user feedback collection, analysis, and triangulation, despite the significant number of tools that have been developed and proposed for user feedback analysis and prioritization, both commercially and research-driven. %\cite{carreno2013analysis, guzman2014users, iacob2013retrieving, maalej_bug_2015, oehri2020same, panichella2015can, gao2022listening, licorish2017attributes, kifetew2021automating, gartner2012method, malgaonkar2022prioritizing}.
Future research could study improving these tools and making them more accessible for practitioners. 

\textbf{Definition and detection of Themes and Trends:} 
Our study highlighted the importance of user feedback themes and the trends that may quickly emerge from the various feedback channels. 
However, how these organizations \emph{define} or \emph{quantify} trends was quite ad-hoc.
Pointing out a trend or theme often relied on an employee's conscience or feeling.
Further research can investigate strategies to better define, quantify or automatically identify themes and trends so that a more systematic and consistent approach may be leveraged by organizations.

\section{Acknowledgements}
This work was supported by the Natural
Sciences and Engineering Research Council of Canada (NSERC).
We are also grateful to each participant who graciously gave us their time and allowed us to interview them.

\bibliographystyle{ACM-Reference-Format}
\bibliography{main.bib}

%\newpage
%\input{supplementaryMaterials.tex}

\end{document}